\documentclass[a4paper]{jpconf}
\usepackage{graphicx}
\begin{document}
\title{Observation of CR Anisotropy with ARGO-YBJ}

\author{Giuseppe Di Sciascio and Roberto Iuppa\\ (on behalf of the ARGO-YBJ Collaboration)}
\address{INFN, Sezione di Roma Tor Vergata, Roma, IT}
\ead{giuseppe.disciascio@roma2.infn.it, roberto.iuppa@roma2.infn.it}

\begin{abstract}

The measurement of the anisotropies of cosmic ray arrival direction provides important informations on the propagation mechanisms and on the identification of their sources.
In this paper we report the observation of anisotropy regions at different angular scales. 
In particular, the observation of a possible anisotropy on scales between $\sim$10$^{\circ}$ and $\sim$30$^{\circ}$ suggests the presence of unknown features of the magnetic fields the charged cosmic rays propagate through, as well as potential contributions of nearby sources to the total flux of cosmic rays. 
Evidence of new weaker few-degree excesses throughout the sky region $195^{\circ}\leq$ R.A. $\leq 315^{\circ}$ is reported for the first time.
\end{abstract}

\section{Introduction}

As cosmic rays (CRs) are mostly charged nuclei, their arrival direction is
deflected and highly isotropized by the action of galactic magnetic field (GMF) they propagate through before reaching the Earth atmosphere. The GMF is the superposition of regular field lines and chaotic contributions. Altough the strength of the non-regular component is still under debate, the local total intensity is supposed to be $B=2\div 4\textrm{ $\mu$G}$. In such a field, the gyroradius of CRs is given by
$r_{a.u.}=100\,R_{\textrm{\scriptsize{TV}}}$,
where $r_{a.u.}$ is in astronomic units  and $R_{\textrm{\scriptsize{TV}}}$ is in TeraVolt.

However, different experiments \cite{nagashima,kam07,tibet06,milagro09,eastop09,icecube11} observed an energy-dependent \emph{"large scale"} anisotropy in the sidereal time frame with an amplitude of about 10$^{-4}$ - 10$^{-3}$, suggesting the existence of two distint broad regions, one showing an excess of CRs (called "tail-in"), distributed around 40$^{\circ}$ to 90$^{\circ}$ in Right Ascension (R.A.). The other a deficit (the "loss cone"), distributed around 150$^{\circ}$ to 240$^{\circ}$ in R.A..
The origin of these anisotropies is still unknown. Some authors claim that it can be explained within the diffusion approximation taking into account the role of the few most nearby and recent sources \cite{blasi,erlykin06}.
Other studies suggest that the observations may be due to a combined effect of the regular and turbolent GMF \cite{battaner09}, or to local uni- and bi-dimensional inflows \cite{amenomori10}.

In the last years the Tibet AS$\gamma$ \cite{amenomori07} and Milagro \cite{milagro08} Collaborations reported evidence of the existence of a medium angular scale anisotropy contained in the tail-in region.
The observation of similar small scale anisotropies has been recently claimed by the Icecube experiment in the Southern hemisphere \cite{icecube11}.
So far, no theory of CRs in the Galaxy exists which is able to explain few degrees anisotropies in the rigidity region 1-10 TV leaving the standard model of CRs and that of the local GMF unchanged at the same time. More beamed the anisotropies and lower their energy, more difficult to fit the standard model of CRs and GMF to experimental results.

The observation of anisotropy effects at a level of 10$^{-4}$ with an air shower array is a tricky job.
A wrong estimation of the exposure may affect the significance and the
relative intensity sky maps, even create artifacts (i.e. fake excesses or
deficit regions).
In fact, drifts in detector response and atmospheric effects on air shower development are quite hard to be modeled to sufficient accuracy.
The envisaged solution is to use the data to estimate the detector exposure. 
But data contain either signal and background events, so that some distortions could be present in the results. 
The shape and the size of possible artifacts depend on the characteristic angle and time scale over which all the aspects of the data acquisition vary less than the effect to be observed.
Therefore, if anisotropies of the order 10$^{-4}$ are looked for,
operating conditions must be kept (or made up) stable down to this level,
all across the field of view and during all the acquisition time.
In addition, as it is well known, the partial sky coverage carried out by a given experiment will bias the observations of structure on largest scales.
Finally, the observation of a possible small angular scale anisotropy region contained inside a larger one rely on the capability for suppressing the anisotropic structures at larger scales without, simultaneuosly, introducing effects of the analysis on smaller scales.

The ARGO-YBJ experiment, located at the YangBaJing Cosmic Ray
Laboratory (Tibet, P.R. China, 4300 m a.s.l., 606 g/cm$^2$), is an air shower array able to detect the cosmic radiation at an energy threshold of a few hundred GeV. The full detector is in stable data taking since November 2007 with a duty cycle greater than 85\%. The trigger rate at the threshold is 3.6 kHz. The detector characteristics and performance are described in \cite{moon11}, the main results are reported in \cite{demitri11}.
In this paper the observation of CR anisotropy at different angular scales with ARGO-YBJ is reported as a function of the primary energy.

\section{Data analysis and Results}

In order to study the anisotropy at different angular scales the isotropic background of CRs has been estimated with two methods: the equi-zenith angle method \cite{amenomori05} and the direct integration method \cite{Fleysher}.
The equi-zenith angle method, used to study the large scale anisotropy, is able to eliminate various spurious effects caused by instrumental and environmental variations, such as changes in pressure and temperature that are hard to control and tend to introduce systematic errors in the measurement.
The direct integration method, based on time-average, rely on the assumption that the local distribution of the incoming CRs is slowly varying and the time-averaged signal may be used as a good estimation of the background content. 
Time-averaging methods act effectively as a high-pass filter, not allowing to inspect features larger than the time over which the background is computed (i.e., 15$^{\circ}$/hour$\times \Delta t$ in R.A.). The time interval used to compute the average spans $\Delta t$= 3 hours and makes us confident the results are reliable for structures up to $\approx$35$^{\circ}$ wide. 
%
\begin{figure}[t!]
  \hspace{-0.8cm}
\begin{minipage}[t]{.5\linewidth}
\begin{center}
 \includegraphics[width=0.85\textwidth]{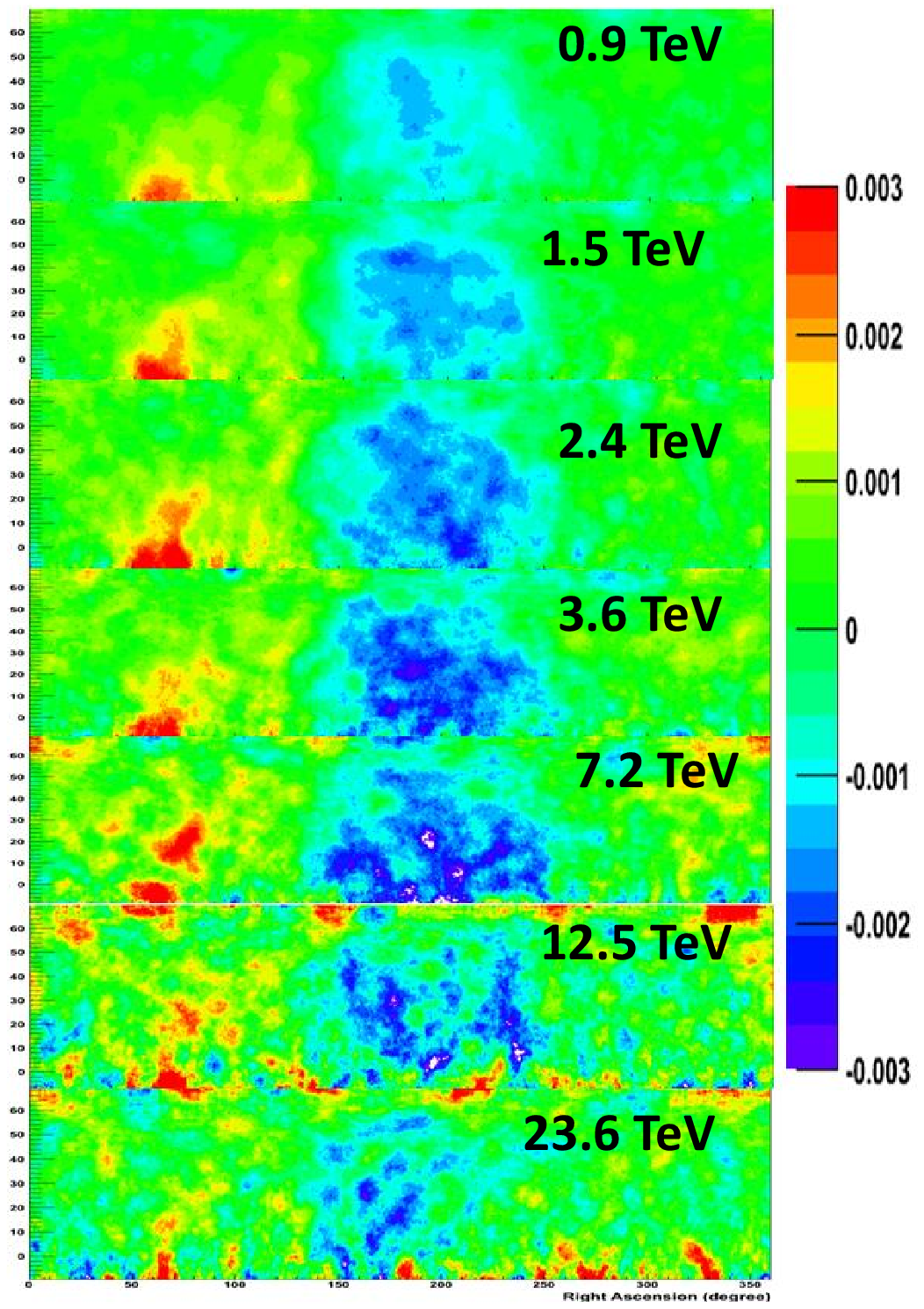}\\
   \end{center}
\end{minipage}\hfill
%
\begin{minipage}[t]{.52\linewidth}
  \begin{center}
  \includegraphics[width=0.95\textwidth]{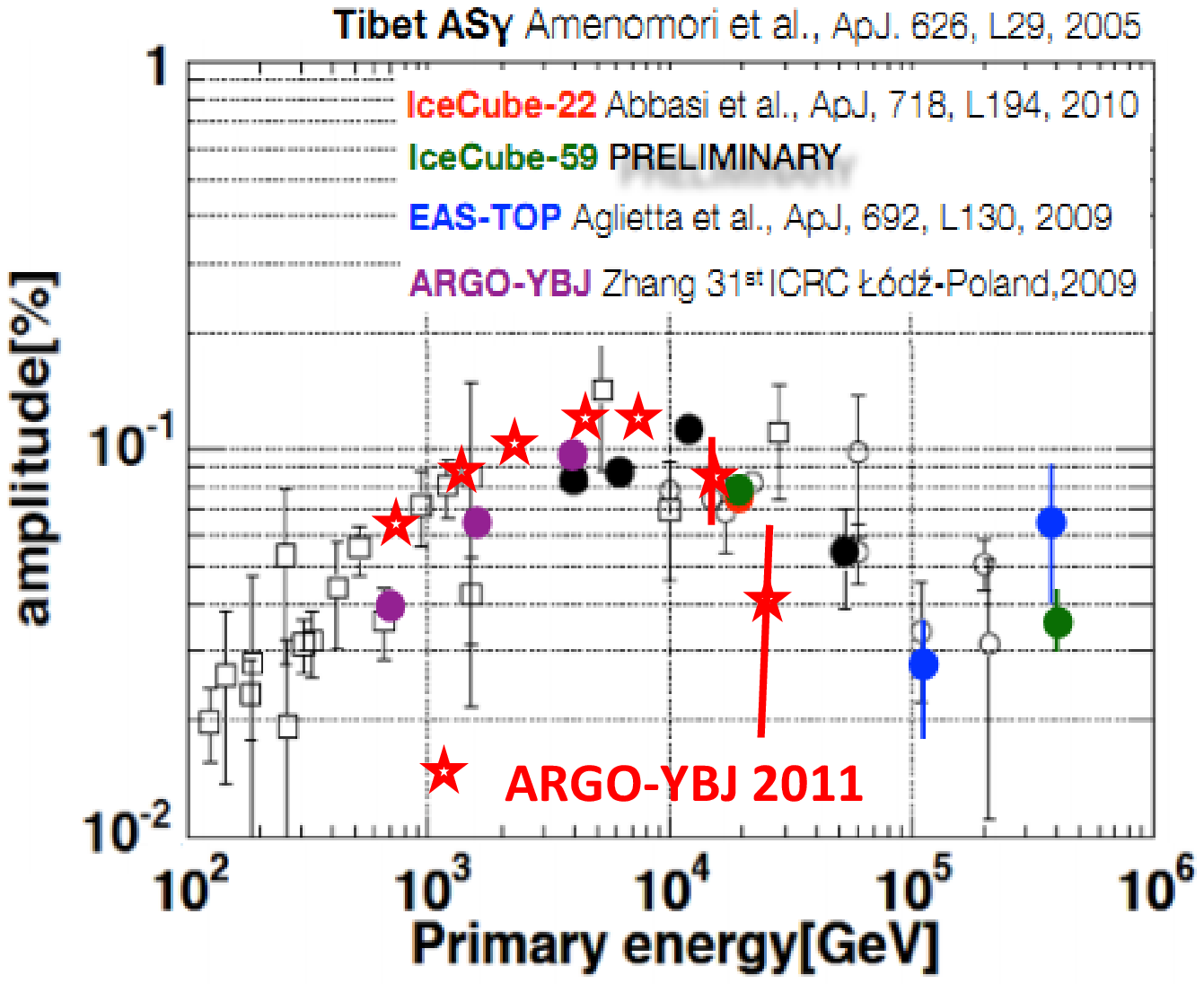}\\
  \end{center}
\end{minipage}\hfill
  \caption{Left plot: Large scale CR anisotropy observed by ARGO-YBJ as a function of the energy. The color scale gives the relative CR intensity.
  Right plot: Amplitude of the first harmonic as a function of the energy, compared to other measurements.} 
\label{fig1}
\end{figure}
%

\subsection{Large Scale Anisotropy}
The observation of the CR large scale anisotropy by ARGO-YBJ is shown in the left plot of Fig. \ref{fig1} as a function of the primary energy up to about 25 TeV. 
The data used in this analysis was collected by ARGO-YBJ from 2008 January
to 2009 December with a reconstructed zenith angle $\leq$ 45$^{\circ}$.
The so-called \textit{`tail-in'} and \textit{`loss-cone'} regions, correlated to an enhancement and a deficit of CRs, are clearly visible with a statistical significance greater than 20 s.d..
The tail-in broad structure appears to dissolve to smaller angular scale spots with increasing energy.
To quantify the scale of the anisotropy we studied the 1-D R.A. projections integrating the sky maps inside a declination band given by the field of view of the detector. Therefore, we fitted the R.A. profiles with the first two harmonics. The resulting amplitude of the first harmonic is plotted in the right plot of Fig. \ref{fig1} where is compared to other measurements as a function of the energy. The ARGO-YBJ results are in agreement with other experiments suggesting a decrease of the anisotropy first harmonic amplitude with increasing energy.
%
\begin{figure}[t!]
  \hspace{-0.8cm}
\begin{minipage}[t]{.5\linewidth}
\begin{center}
 \includegraphics[width=0.95\textwidth]{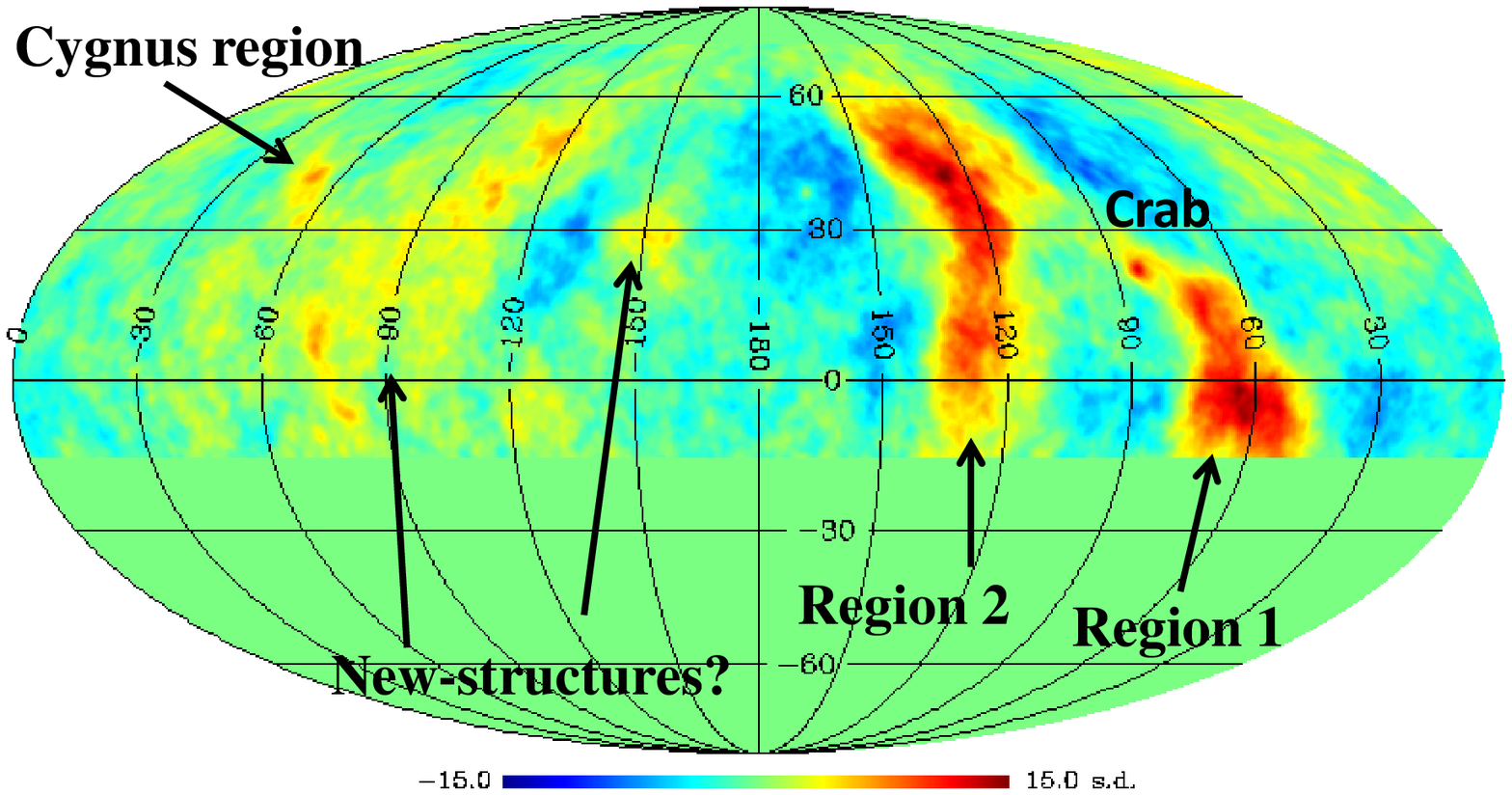}\\
   \end{center}
\end{minipage}\hfill
%
\begin{minipage}[t]{.52\linewidth}
  \begin{center}
  \includegraphics[width=0.85\textwidth]{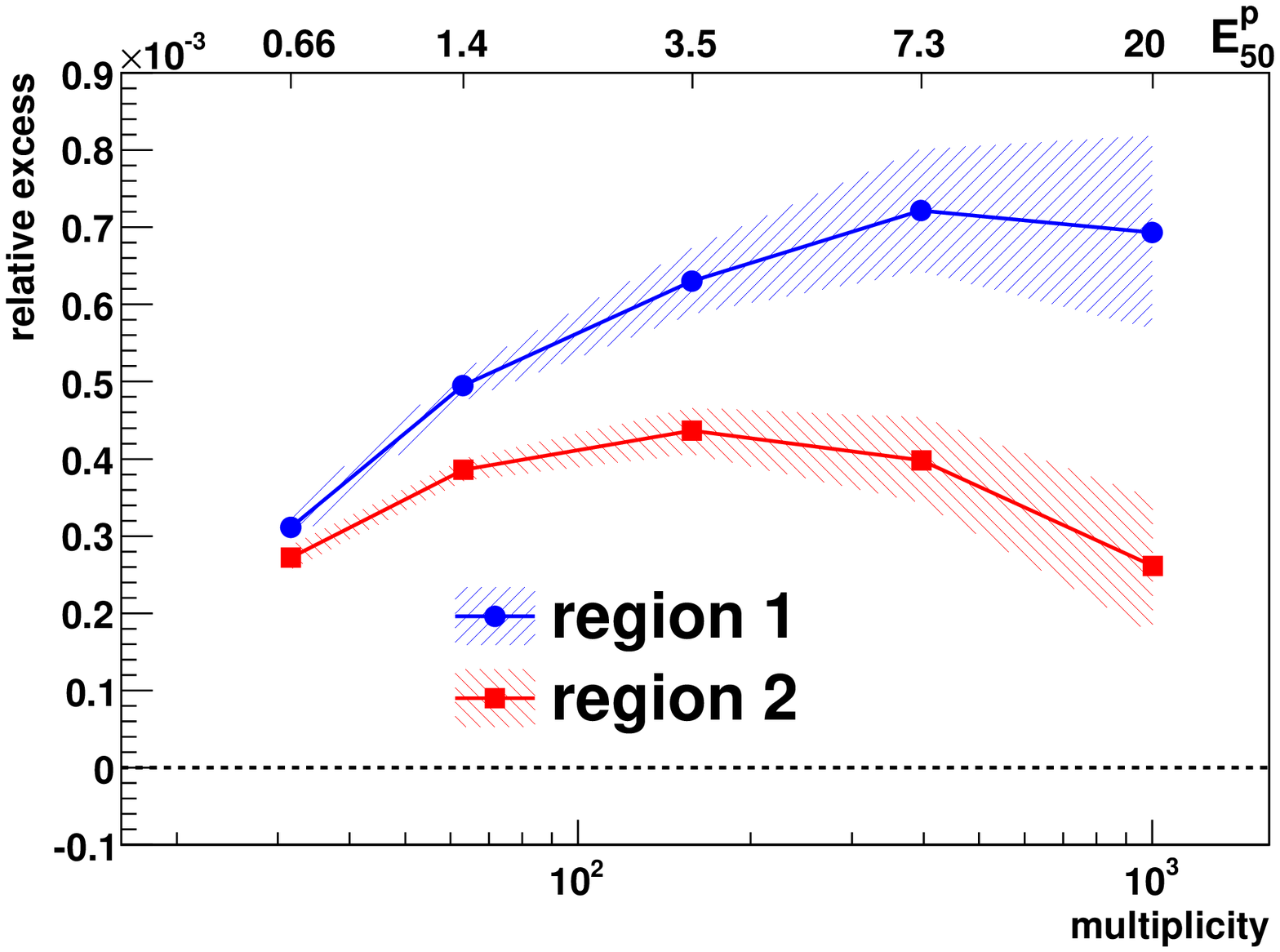}\\
  \end{center}
\end{minipage}\hfill
  \caption{Left plot: Intermediate scale CR anisotropy observed by ARGO-YBJ. The color scale gives the statistical significance of the observation in standard deviations. 
  Right plot: Size spectrum of the regions 1 and 2. The vertical axis represents the ratio between the events collected. The upper scale shows the corresponding proton median energy (see text).} 
\label{fig2}
\end{figure}
%

\subsection{Intermediate Scale Anisotropy}
The left plot of the Fig. \ref{fig2} shows the ARGO-YBJ sky map in equatorial coordinates.
The analysis refers to events collected from November 2007 to May 2011 after the following selections: (1) $\geq$25 shower particles on the detector; (2) zenith angle of the reconstructed showers $\leq$50$^{\circ}$.
The triggering showers that passed the selection were about 2$\cdot$10$^{11}$. The zenith cut selects the declination region $\delta\sim$ -20$^{\circ}\div$ 80$^{\circ}$.
According to the simulation, the median energy of the isotropic cosmic ray proton flux is E$_p^{50}\approx$1.8 TeV (mode energy $\approx$0.7 TeV).

The most evident features are observed by ARGO-YBJ around the positions $\alpha\sim$ 120$^{\circ}$, $\delta\sim$ 40$^{\circ}$ and $\alpha\sim$ 60$^{\circ}$, $\delta\sim$ -5$^{\circ}$, positionally coincident with the regions detected by Milagro \cite{milagro08}. These regions, named ``region 1'' and ``region 2'', are observed with a statistical significance of about 14 s.d.. 
The deficit regions parallel to the excesses are due to a known effect of the analysis, that uses also the excess events to evaluate the background, artificially increasing the background.
On the left side of the sky map, several possible new extended features are visible, though less intense than those aforementioned. The area $195^{\circ}\leq R.A. \leq 315^{\circ}$ seems to be full of few-degree excesses not compatible with random fluctuations (the statistical significance is more than 6 s.d. post-trial). 
The observation of these structures is reported here for the first time and together with that of regions 1 and 2 it may open the way to an interesting study of the TeV CR sky.
To figure out the energy spectrum of the excesses, data have been divided into five independent shower multiplicity sets. The number of events collected within each region are computed for the event map as well as for the background one. The ratio of these quantities is computed for each multiplicity interval. The result is shown in the right plot of the Fig. \ref{fig2}. Region 1 seems to have spectrum harder than isotropic CRs and a cutoff around 600 shower particles (proton median energy E$^{50}_p$ = 8 TeV). On the other hand, the excess hosted in region 2 is less intense and seems to have a spectrum more similar to that of isotropic cosmic rays.
The steepening from 100 shower particles on (E$_p^{50}$ = 2 TeV) is
likely related to efficiency effects. Further studies are on the way.

\section{Conclusions}
In this paper the observation of CR anisotropy at different angular scales with ARGO-YBJ is reported as a function of the primary energy.
The large scale CR anisotropy has been clearly observed up to about 25 TeV.
Evidence of existence of different few-degree excesses in the Northern sky (the strongest ones positionally coincident with the regions detected by Milagro in 2008) is reported.
The study of the CR anisotropy with air shower arrays is challenging, therefore deeper analysis is under way to investigate possible artifacts produced in the background calculations.
However, a joint analysis of concurrent data recorded by different experiments in both hemispheres, as well as a correlation with other observables like the interstellar energetic neutral atoms distribution \cite{ibex09,ibex11}, should be a high priority to clarify the observations.

\smallskip

\end{document}